\documentclass[aps,onecolumn,showpacs,preprint,endfloats,amssymb]{revtex4}
\makeatletter
\def\@dotsep{4.5}
\makeatother
\usepackage{graphicx}

\selectfont

\begin{document}

\title{Bifurcations as dissociation mechanism in bichromatically driven diatomic molecules}

\author{S. Huang$^1$}
\email{gtg098n@mail.gatech.edu}
\author{C. Chandre$^2$}
\author{T. Uzer$^1$}

\affiliation{$^1$ Center for Nonlinear Science, School of Physics,
Georgia Institute of Technology, Atlanta, Georgia 30332-0430, U.S.A.\\
$^2$ Centre de Physique Th\'eorique\footnote{UMR 6207 of the CNRS,
Aix-Marseille and Sud Toulon-Var Universities. Affiliated with the
CNRS Research Federation FRUMAM (FR 2291). CEA registered research
laboratory LRC DSM-06-35.}, Luminy - case 907, 13288 Marseille
cedex 09, France}

\date{\today}

\begin{abstract}
We discuss the influence of periodic orbits on the dissociation of
a model diatomic molecule driven by a strong bichromatic laser fields.
Through the stability of periodic orbits we analyze the
dissociation probability when parameters like the two amplitudes
and the phase lag between the laser fields, are varied. We find that
qualitative features of dissociation can be reproduced by
considering a small set of short periodic orbits. The good agreement
with direct simulations demonstrates the importance of
bifurcations of short periodic orbits in the dissociation
dynamics of diatomic molecules.
\end{abstract}

\pacs{82.50.Pt, 82.50.Nd, 82.20.Nk, 05.45.Gg}

\maketitle

\section{Introduction}
\label{sec1}

The dissociation behavior of molecules driven by bichromatic
fields with commensurate frequencies has emerged as a rich
research subject, especially for the control of molecular
processes \cite{Shapiro,Gordon}.  The interplay of the two
radiation fields opens up many new dissociation pathways, and it
is well known that the relative phase between the two fields can
affect these pathways drastically, so much so that the relative
phase can be used as a means to control the outcome of the
reaction \cite{Charron,batista}. However, the mechanisms by which
the relative phase controls the dissociation behavior are less
well-known. The relative phase is a very convenient control
parameter since it does not require additional energy input from
the fields (as opposed to their amplitudes).

The two-color laser-driven dissociation of molecules is also of great
interest to researchers because these seemingly simple
systems display complex dynamics and behavior that single
component laser field cannot exhibit
\cite{Shapiro,Gordon,Charron,batista,Goggin2,Stine,greeks,Levesque,He}.
The theoretical literature on laser-driven dissociation of molecules has been extensive in the past three decades
\cite{Charron,batista,Goggin2,Stine,greeks,Levesque,He,gu,greeks2,greeks3,Goggin,Guldberg,Wu,Thachuk,Nico,Dardi,Heather,Bandrauk}.
Among these is Ref.~\cite{greeks} where the dissociation probability of a model diatomic
molecule exposed to a two-color laser field was
investigated for various parameters using direct simulations of
the classical-mechanical equations.

In this paper, we report how the dissociation probability
obtained in Ref.~\cite{greeks} by direct numerical simulations can
be understood qualitatively using a linear stability analysis of a
small set of periodic orbits. Our main result is that for most parameter
values the principal features of the dissociation
probability can be reproduced using two short periodic orbits
(with the period equal to the one of the fields), and in
particular by the identification of the main bifurcations which have a drastic effect on the dissociation probability. In
this way, our approach allows the qualitative prediction, with
significant time savings, of the dynamics as parameters are
varied. Indeed, the typical time necessary for the computation of
a periodic orbit and its stability is of the order of its period
which is also the period of the field. Our findings echo similar
ones obtained in the microwave ionization of Rydberg atoms in a
strong bichromatic field for which qualitative agreement has
been obtained with experimental data and a quantitative agreement
with quantal simulations~\cite{huang}.

After describing the model in Sec.~\ref{sec0}, we briefly
summarize the method which monitors the position and the residue
of periodic orbits \cite{bach06} in Sec.~\ref{sec2}. In
Sec.~\ref{sec3}, we analyze the dynamics using a selection of
short periodic orbits, as parameters are varied in
Sec.~\ref{sec:IIIA}. We relate a linear stability measure (the
residue of a given periodic orbit) to the dissociation probability
in Sec.~\ref{sec:IIIB}. Good agreement is found for most parameter values. A discrepancy is observed for small
phases and is discussed in Sec.~\ref{sec:IIIC}.

\section{The model}
\label{sec0}

The Hamiltonian (in atomic units) of a diatomic molecule exposed to a strong
bichromatic field with a phase lag can be modeled as
\begin{eqnarray}
 &&H(r,p,\tau)=\frac{p^2}{2m}+D\left[1-{\rm e}^{-\alpha(r-r_e)}\right]^2\nonumber \\
 &&\quad+(r-r_e)[A_1 \sin(\Omega_1 \tau)+A_2 \sin(\Omega_2 \tau+\phi)],\label{Molecule}
\end{eqnarray}
where the parameters are the reduced mass $m$, the dissociation
energy $D$, and the equilibrium distance $r_e$. Here we assume
that the envelopes of the pulses are constant since the pulse
duration has a very minor impact on this system, as suggested in
Ref.~\cite{greeks}. The relevant dimensionless variables are
$\tilde{r}=\alpha(r-r_e)$, $\tilde{p}=p/\sqrt{2Dm}$,
$t=\alpha\sqrt{2D/m}\tau$, $F_{i}=A_{i}/(2D\alpha)$,
$\omega_{i}=\Omega_{i}/\sqrt{2D\alpha^{2}/m}$. In these new
coordinates, the Hamiltonian~(\ref{Molecule}) is
\begin{eqnarray}
\tilde{H}(\tilde{r},\tilde{p},t)&=&\frac{\tilde{p}^2}{2}+\frac{1}{2}\left(1-{\rm e}^{-\tilde{r}}\right)^2\nonumber \\
&&+\tilde{r}\left(F_1\sin \omega_1 t
+F_2\sin(\omega_2t+\phi)\right),\label{eqn:Hdim}
\end{eqnarray}
where $\tilde{r}$ and $\tilde{p}$ are canonically conjugate.

In what follows, we model the hydrogen fluoride (HF) molecule
in which $m=1732$, $D=0.2101$, $r_{e}=1.75$ and $\alpha=1.22$ (all in
atomic units.) We consider a field with two commensurate
frequencies such that $\omega_1=\omega_{2}/3=0.28$. We notice that
Hamiltonian~(\ref{eqn:Hdim}) is time-periodic with period
$2\pi/\omega_1$.

\section{Residue method}
\label{sec2} The general idea of the residue method is to follow a
set of periodic orbits as parameters are varied in order to
determine qualitative properties of the dynamics. As it was shown
in other, similar problems, short periodic orbits play the role of
organizing centers for the dynamics~\cite{Joyeux,Farantos,huang}.
Higher-order periodic orbits give more refined details of the
dynamics, especially on longer time scales. In principle,
for atomic and molecular systems where short pulses are
considered, only short periodic orbits should influence the
dynamics. We determine the location of a periodic orbit (given by
its number of intersections with the apt Poincar\'e surface of
section) using a modified Newton-Raphson multi-shooting algorithm
as described in Ref.~\cite{chaosbook}. The initial conditions from
which the Newton map are iterated is determined in two possible
ways~: By a quick inspection of the Poincar\'e section (which is
easier if the periodic orbit is elliptic since it shows resonant
islands around the periodic orbit considered) or by continuation
of periodic orbits for other values of parameters (which is
clearly the optimal way since periodic orbits usually deform
continuously as parameters are varied, and even bifurcate). We
also monitor the linear stability properties of these periodic
orbits which are obtained by integrating the reduced tangent flow
along the periodic orbit
$$
\frac{d{\mathcal J}^t}{dt}={\mathbb J}\nabla^2
\tilde{H}(\tilde{r},\tilde{p},t) {\mathcal J}^t,
$$
where ${\mathbb J}=\left(\begin{array}{cc} 0 & 1\\ -1 & 0
\end{array}\right)$ and $\nabla^2 \tilde{H}$ is the two-dimensional
Hessian matrix (composed of second derivatives of $\tilde{H}$ with
respect to its canonical variables $\tilde{r}$ and $\tilde{p}$).
The initial condition is ${\mathcal J}^0={\mathbb I}_2$ (the
two-dimensional identity matrix). For a periodic orbit with period
$T$ where $T=2\pi n /\omega_1$ ($n$ being the number of
intersections with the Poincar\'e section), the two eigenvalues of
the monodromy matrix ${\mathcal J}^{T}$ ( which make the pair
$(\lambda,1/\lambda)$) determine the stability properties. The
determinant of ${\mathcal J}^{T}$ is equal to 1 since the flow is
volume preserving. If the spectrum is $({\mathrm
e}^{i\omega},{\mathrm e}^{-i\omega})$, the periodic orbit is
elliptic (stable, except in some rare cases); or hyperbolic if
the spectrum is $(\lambda,1/\lambda)$ with $\lambda\in{\mathbb
R}^*$ (unstable). Through the use of Greene's residue
$R$~\cite{gree79,mack92}
$$
R=\frac{2-\mbox{tr}{\mathcal J}^{T}}{4},
$$
the stability properties can be deduced in a concise form. If
$R\in ]0,1[$, the periodic orbit is elliptic; if $R<0$ or $R>1$ it
is hyperbolic; and if $R=0$ and $R=1$, it is parabolic.

For a given periodic orbit, we follow its location in phase space
and its residue as parameters are varied. There are
three parameters in our problem: Two amplitudes $F_1$ and $F_2$, and a phase lag
$\phi$. We compute $R(F_1,F_2,\phi)$ and identify the points in
parameter space where bifurcations occur. We are interested in the
bifurcations whenever a periodic orbit is likely to change its
linear stability, which occurs in particular for
$R(F_1,F_2,\phi)=0$ or $R(F_1,F_2,\phi)=1$. In general such
bifurcations (based on a linear stability analysis) will also play
an important role in the nearby phase space region (by continuity
in phase space).

\section{Dissociation probability}
\label{sec3}

\subsection{Identification of fundamental periodic orbits}
\label{sec:IIIA}

\begin{figure}
 \centering
 \includegraphics[width=12.5cm,height=11.5cm]{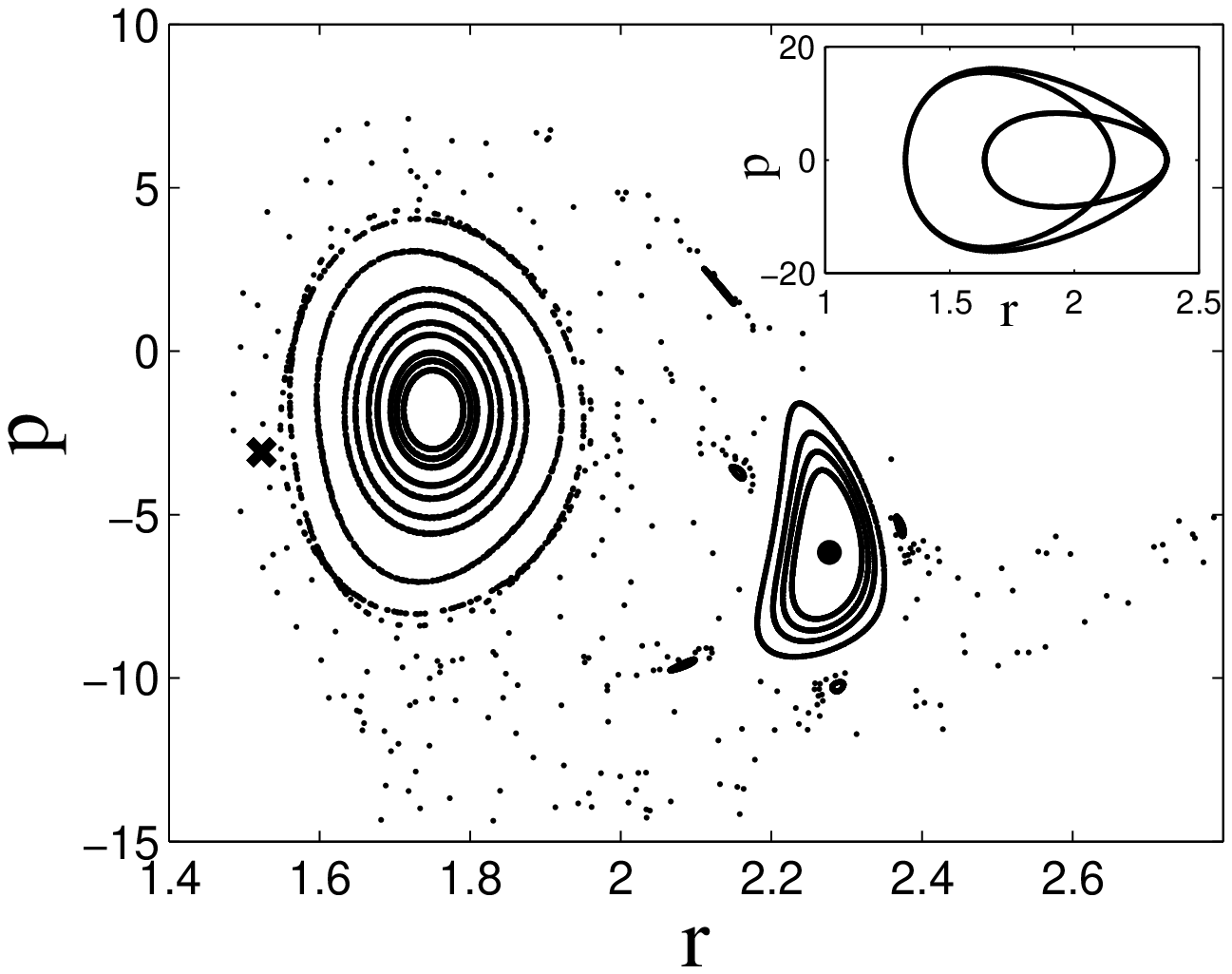}
\caption{\label{fig:fig1}Poincar\'e section of
Hamiltonian~(\ref{eqn:Hdim}) for $F_1=0.18$, $F_2=0.02$ and
$\phi=0$. The dot and the cross indicate the elliptic periodic
orbit ${\mathcal O}_e$ and its associated hyperbolic one,
respectively. The inset depicts a projection of ${\mathcal O}_e$
in the $(r,p)$ plane.}
\end{figure}

Figure~\ref{fig:fig1} shows a Poincar\'e section (stroboscopic
plot of phase space with period $2\pi/\omega_1$) of
Hamiltonian~(\ref{eqn:Hdim}) for amplitudes $F_1=0.18$, $F_2=0.02$
and phase lag $\phi=0$. We notice that an elliptic island is
present at the entrance of the dissociation channel. At the center
of this island sits an elliptic periodic orbit with one
intersection with the Poincar\'e surface of section (i.e.\ with
period $2\pi/\omega_1$). Standard Hamiltonian dynamics show that
the trajectories that are likely to dissociate can become trapped
around the resonant island for a while before finding an escape
route. Therefore, this particular periodic orbit, which we call
${\mathcal O}_e$, plays a crucial role in the dissociation
probability, and is the focus of the current paper~: We
investigate its role as the parameters $(F_1,F_2,\phi)$ are
varied. Due to the symmetry ($\phi\mapsto \pi-\phi$) the
fundamental domain of variations of $\phi$ is $[0,\pi[$. We also
restrict the amplitudes to $(F_1,F_2)\in[0,0.22]\times[0,0.06]$.

\begin{figure}
 \centering
 \includegraphics[width=12.5cm,height=11.5cm]{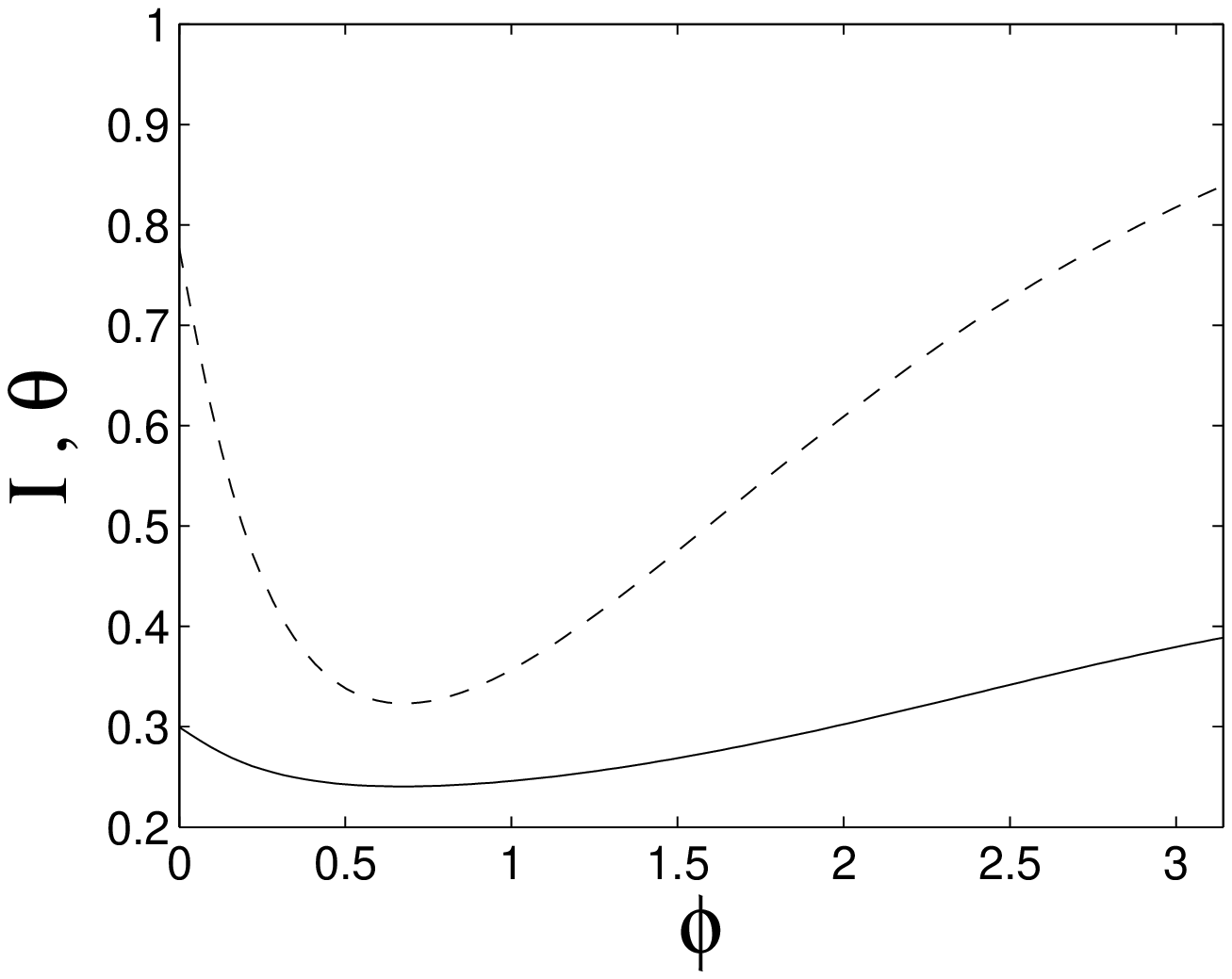}
\caption{\label{fig:position} Action(solid) and angle(dashed)
curves for $F_1=0.18$ and $F_2=0.02$.}
\end{figure}

We anticipate two factors which can influence dissociation: One
is the location of this orbit, and the other is a change of
its stability. In Fig.~\ref{fig:position}, we represented the
position of ${\mathcal O}_e$ or, more precisely, its action and
angle variables as defined by~\cite{gu}
\begin{eqnarray*}
    && I=2\left( 1-\sqrt{1-E}\right),\\
    && \tan \theta=-\frac{\tilde{p}\sqrt{1-E}}{1-{\rm e}^{-\tilde{r}}-E},
\end{eqnarray*}
where $E(\tilde{r},\tilde{p})=\tilde{p}^2+(1-{\rm
e}^{-\tilde{r}})^2$, for a typical set of parameters $F_1=0.18$
and $F_2=0.02$, while $\phi$ is varied. The main conclusion is
that the variation of the action and angle of ${\mathcal O}_e$ do
not seem to be linked to the variation of the dissociation
probability since they do not vary significantly as $\phi$ is
increased. In addition they are not monotonic functions of $\phi$
in contrast with the dissociation probability. Hence, the position
of the specific periodic orbit does not appear to play a
significant role (at least in the range of parameters considered).

\begin{figure}
 \centering
 \includegraphics[width=12.5cm,height=11.5cm]{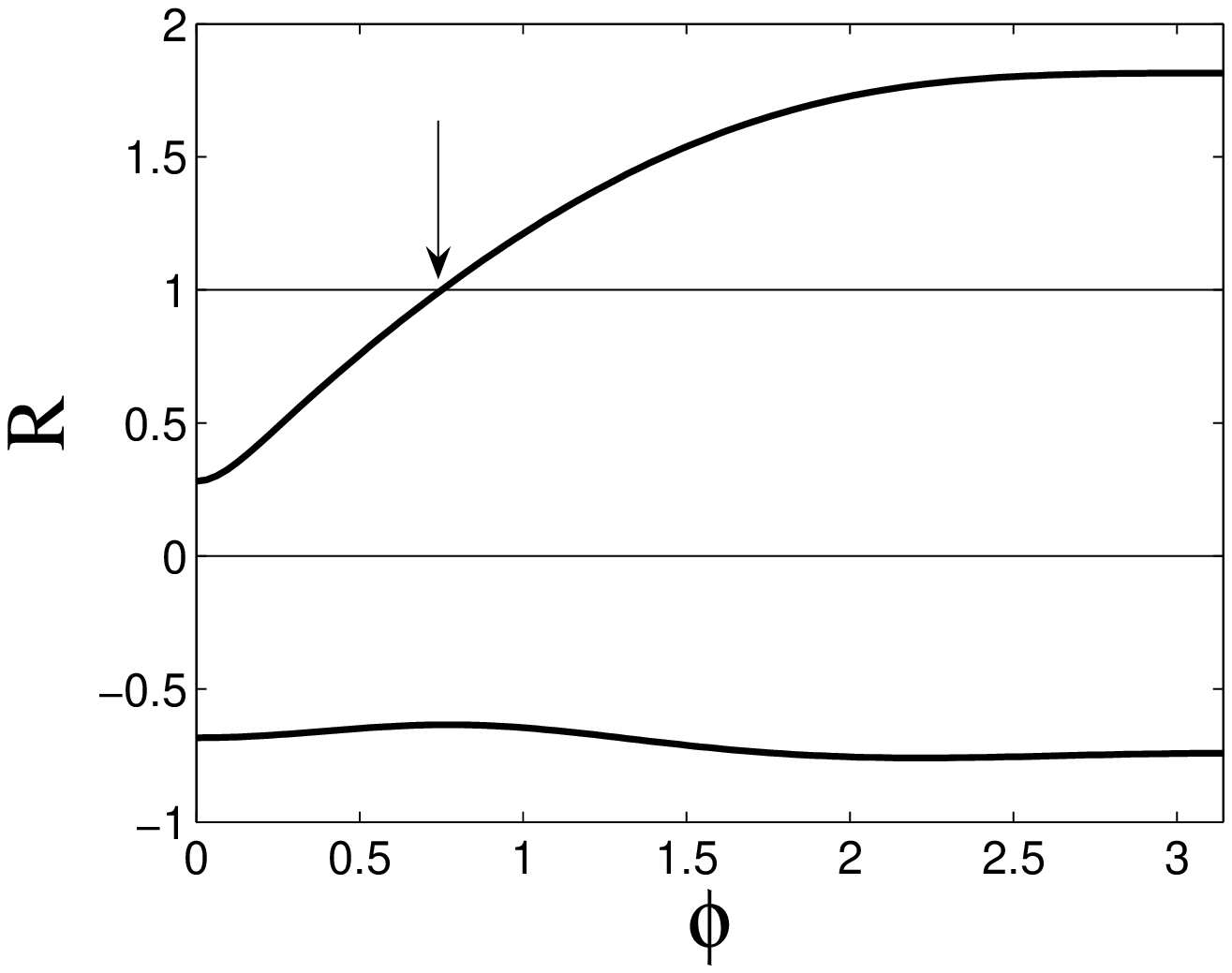}
\caption{\label{fig:residue}Residue curves (bold solid) for the
considered set of periodic orbits for $F_1=0.18$ and $F_2=0.02$~:
for ${\mathcal O}_e$ (upper curve) and for ${\mathcal O}_h$ (lower
curve). The arrow indicates where the bifurcation happens.}
\end{figure}

The second possible mechanism based on a bifurcation has a more
drastic influence on dissociation. In order to monitor
this bifurcation properly, we also need to follow the associated hyperbolic
orbit~\cite{bach06}. To see an increase of dissociation
we need to ensure that the hyperbolic orbit stays hyperbolic while
the elliptic one turns hyperbolic (in order to discard a stability
exchange which would not affect the dissociation
probability significantly). A typical residue plot (the residue as a function of
$\phi$) is shown in Fig.~\ref{fig:residue} for $F_1=0.18$
and $F_2=0.02$. We notice that an increase of $\phi$ is always
associated with an increase of the residue. At $\phi=0.75$, the
residue crosses unity and a bifurcation (which is a period
doubling) occurs. This increase of hyperbolicity is associated
with increased chaos and hence more dissociation. This is in
agreement with the direct simulations of Ref.~\cite{greeks}.

We should point out that the computation of Ref.~\cite{greeks}
uses initial conditions in the ground state ($E=0.045$), which is
lower than the set of periodic orbits we consider ($E=0.29$) in
Fig.~\ref{fig:fig1}. This justifies the importance of the chosen
periodic orbits for dissociating trajectories. For other values of
parameters we consider, the energy level of the periodic orbit
${\mathcal O}_e$ is always well above $E=0.045$.

\subsection{Residue contour plots in parameter space}
\label{sec:IIIB}

\begin{figure}
 \begin{minipage}[t]{8cm}
 \centering
 \includegraphics[width=7cm,height=5.4cm]{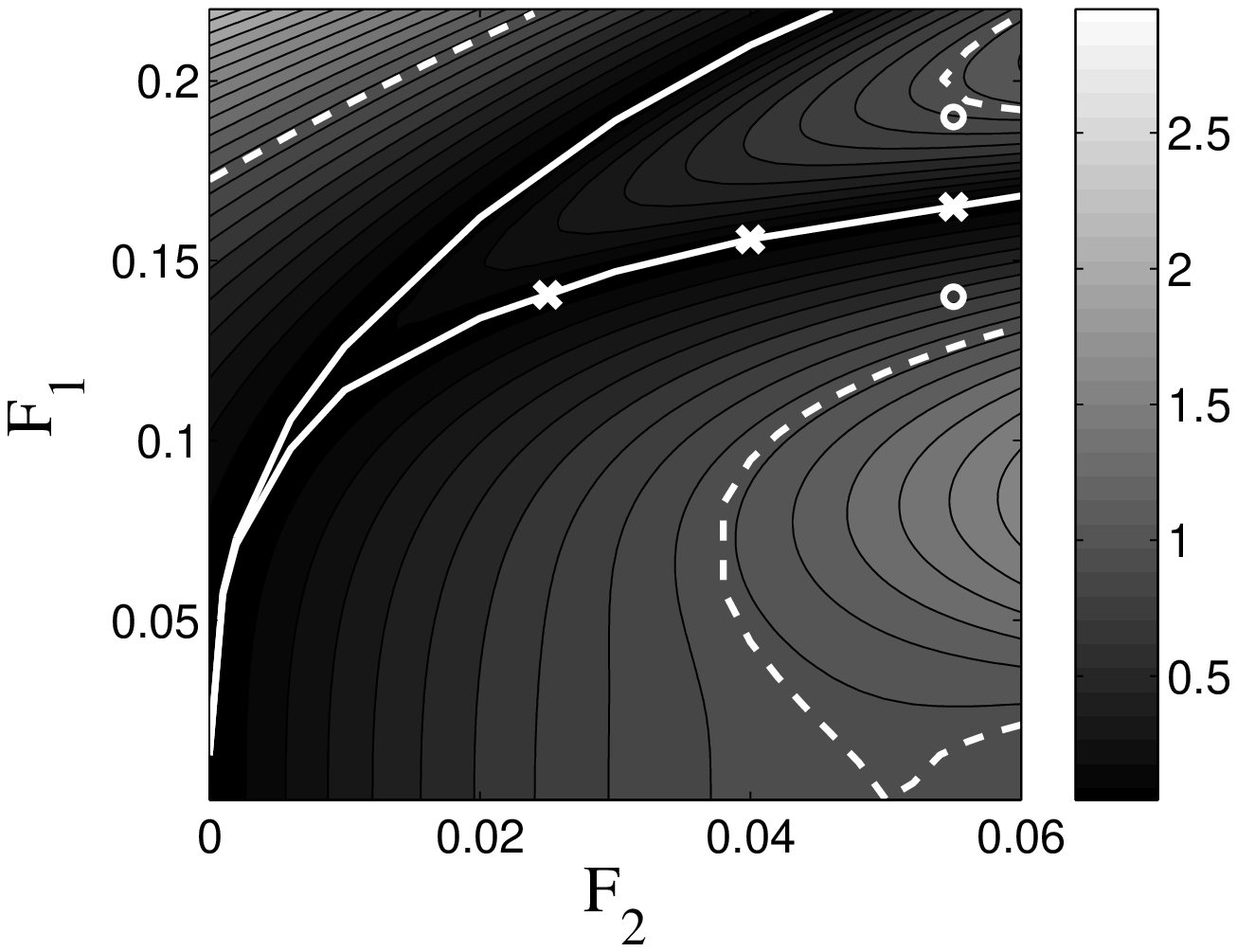}
 \mbox{{\bf (a)} }
 \end{minipage}
 \begin{minipage}[t]{8cm}
 \centering
 \includegraphics[width=7cm,height=5.4cm]{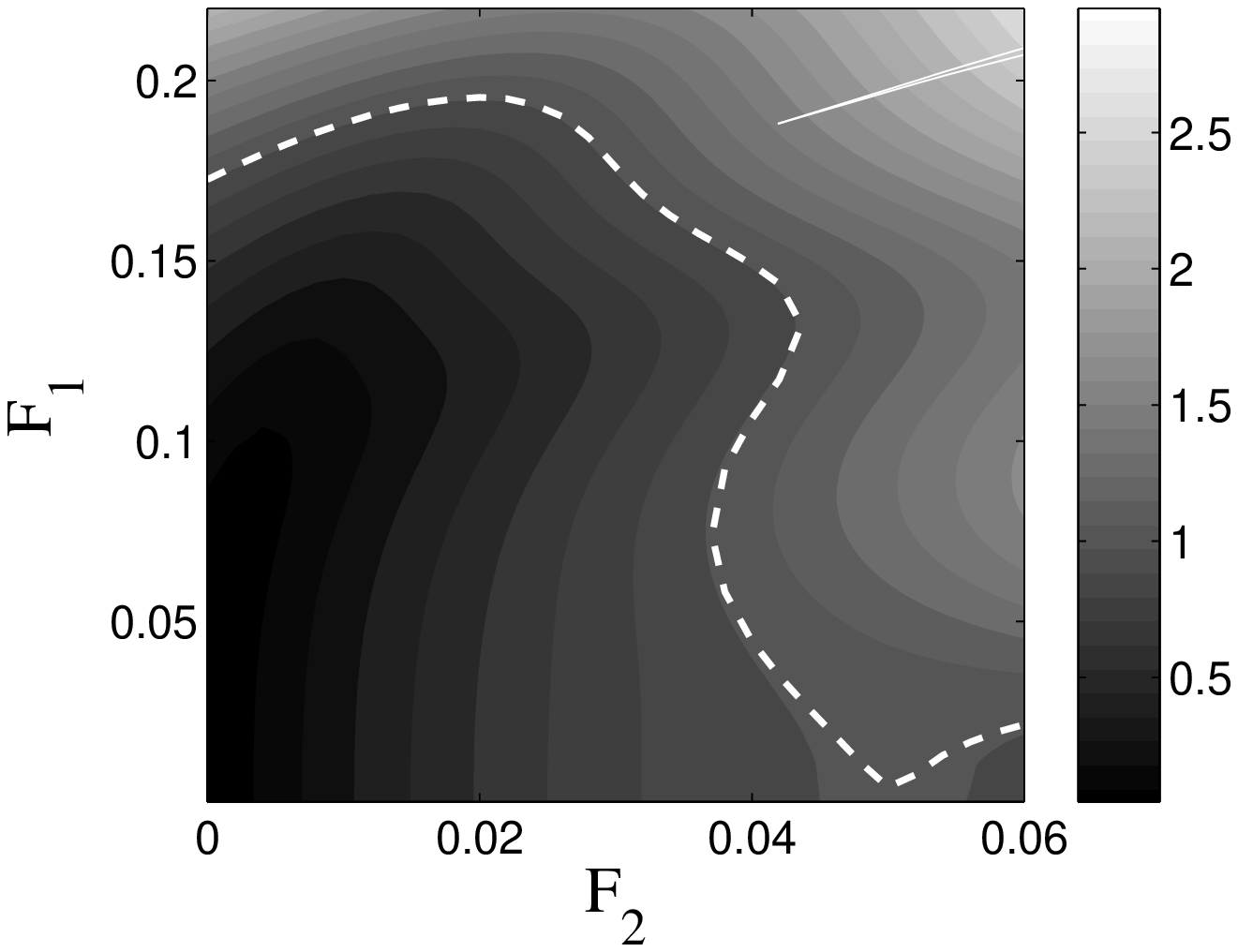}
 \mbox{{\bf (b)} }
 \end{minipage}
 \centering
 \begin{minipage}[t]{8cm}
 \includegraphics[width=7cm,height=5.4cm]{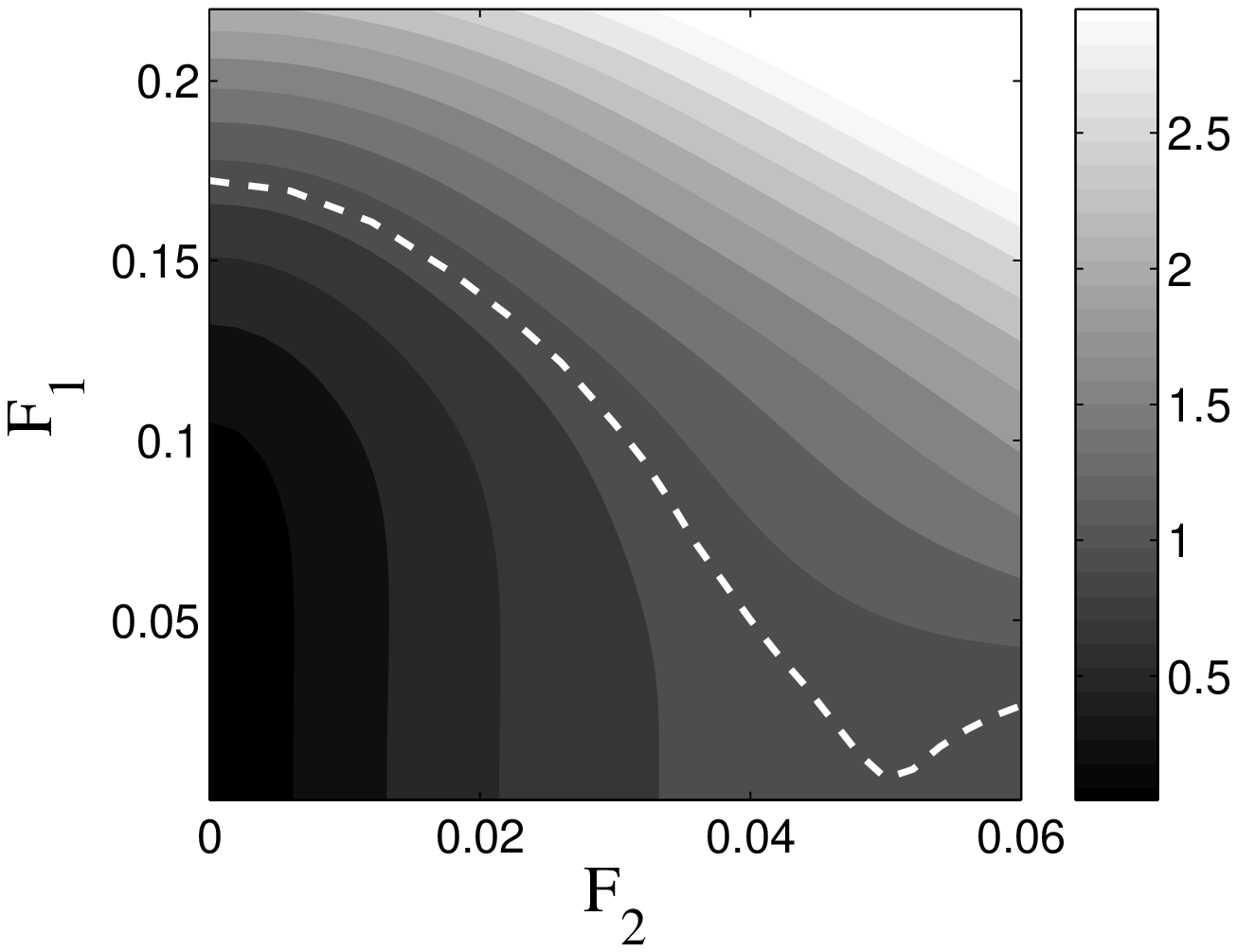}
 \mbox{{\bf (c)} }
 \end{minipage}
 \begin{minipage}[t]{8cm}
 \includegraphics[width=7cm,height=5.4cm]{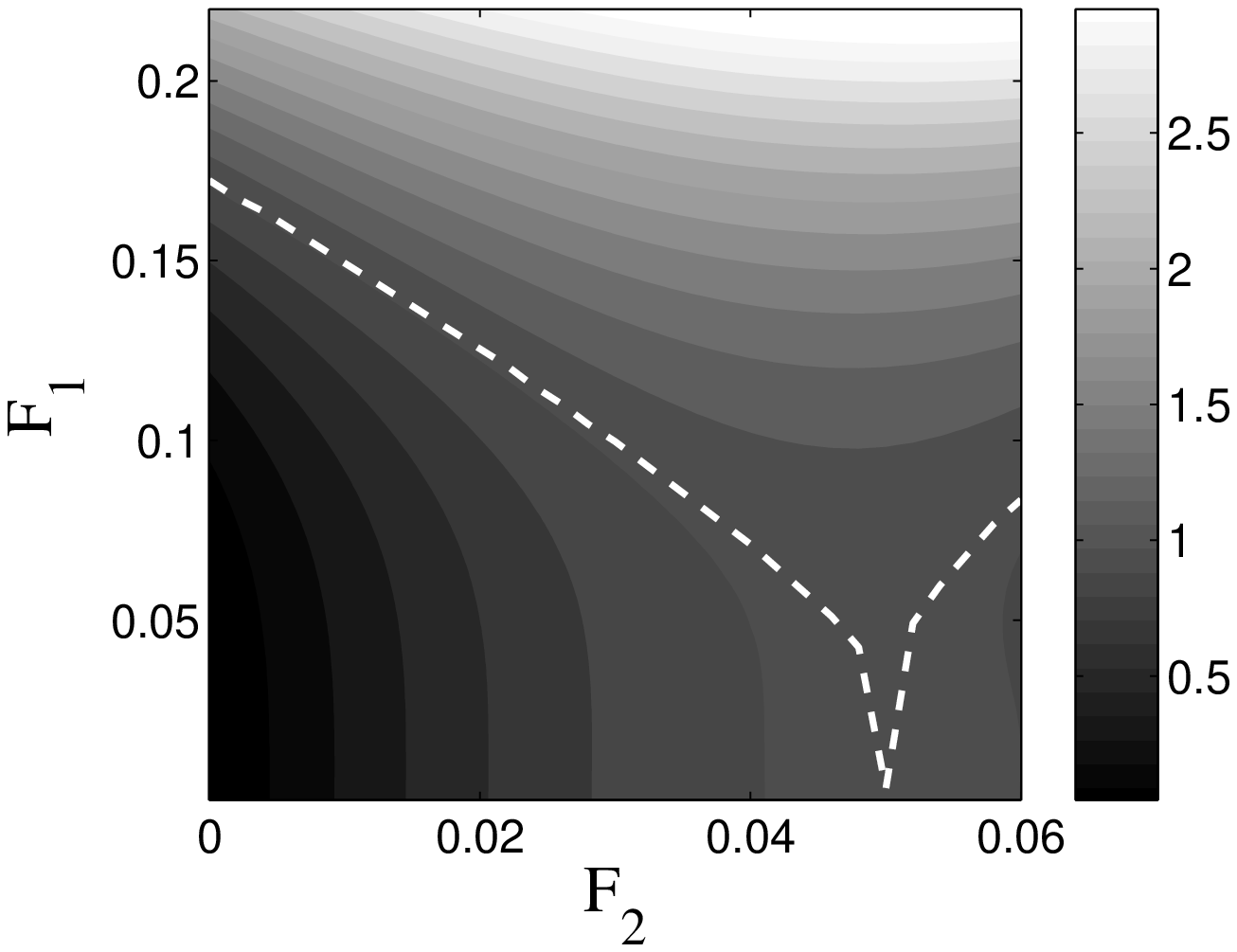}
 \mbox{{\bf (d)} }
 \end{minipage}
 \caption {\label{fig:fig2} Contour plots of the
residues of ${\mathcal O}_e$ in the $(F_1,F_2)$ plane for $(a)$
$\phi=0$, $(b)$ $\phi=\pi/6$, $(c)$ $\phi=\pi/2$ and $(d)$
$\phi=\pi$. White dashed curves indicate $R(F_1,F_2)=1$ where
${\mathcal O}_e$ bifurcates. Two white solid curves in $(a)$
indicate $R(F_1,F_2)=0$. The small region enclosed by two thin
white lines near the up-right corner in $(b)$ corresponds to a
region where ${\mathcal O}_h$ is elliptic. Crosses (respectively circles) on panel $(a)$ represent the values of parameters where laminar plots have been computed in Fig.~\ref{fig:laminaronline} (respectively Fig.~\ref{fig:laminarlocal}).}
\end{figure}

For fixed values of $\phi$, we vary the amplitudes $F_1$ and $F_2$
and compute the residue values of the periodic orbit ${\mathcal
O}_e$. We depict the contour plots of these residues in
Fig.~\ref{fig:fig2} in the $(F_1,F_2)$ plane for $\phi=0$,
$\pi/6$, $\pi/2$ and $\pi$. These figures should be compared to
Fig.~2 of Ref.~\cite{greeks}. In Fig.~\ref{fig:fig2}, darker
regions represent lower residue values, and hence are expected to
reveal more stable dynamics ( smaller dissociation probability)
for the corresponding parameters.

On these residue contours we have superposed the curves
$R(F_1,F_2)=1$ (white dashed curves) through which the periodic
orbit ${\mathcal O}_e$ changes its stability. The corresponding
bifurcation from elliptic to hyperbolic linear stability indicates
a possible increase in dissociation probability due to an increase
of hyperbolicity in this region of phase space. Since the
trajectories are no longer trapped when ${\mathcal O}_e$ is
hyperbolic, it is expected that the dissociation probability
increases. This is in good agreement with Ref.~\cite{greeks} where
we notice that for $\phi$=$\pi/6$, $\pi/2$ and $\pi$, there is a
qualitative agreement in the shape of the dissociation
probability. In particular, the following features are reproduced
for a fixed value of $\phi$~: the non-monotonicity for
$\phi=\pi/6$ as $F_2$ is increased (with a fixed value of $F_1$),
and the monotonicity for $\phi=\pi/2$ and $\pi$ with a sharper
downward bifurcation curve (dashed line) for $\phi=\pi$ as $F_2$
is increased in the region $F_2\in[0, 0.05]$. Our analysis also
confirms that the stabilization effect decreases when $\phi$ is
increased (from 0 to $\pi$). For $\phi=0$ case, the contour plot
shows agreement with direct simulations for most regions in the
$(F_1,F_2)$ plane once more, and the property that it reproduces
the two upper-right bumps observed in the dissociation probability
contour plot. These are interpreted as remnants of the ellipticity
of the bifurcated ${\mathcal O}_e$. We notice that the
corresponding hyperbolic periodic orbit ${\mathcal O}_h$ remains
hyperbolic ($R<0$) for most values of the parameters. However, for
$\phi\in[\pi/6,\pi]$, there is a region around the upper-right
corner of the $(F_1,F_2)$ plane where the ${\mathcal O}_h$ turns
to elliptic and then returns to hyperbolic, as it is shown on the
$(F_1,F_2)$ plane contour plot of Fig.~\ref{fig:fig2} $(b)$. In
general, this bifurcation does not affect the dissociation
probability because, due to its location, it does
not play an important role compared to the periodic orbit
${\mathcal O}_e$ which has already bifurcated ($R>1$) for these parameter
values (see Figs.~\ref{fig:fig2} $(c)$ and $(d)$).
However, for low values of $\phi$ and high values of the
amplitudes $F_i$, the orbit ${\mathcal O}_e$ is still elliptic and
${\mathcal O}_h$ undergoes a bifurcation, as shown in
Figs.~\ref{fig:fig4} which we discuss in the following section.
This region corresponds to the disagreement observed for $\phi=0$
on Fig.~\ref{fig:fig2} with the direct simulations of
Ref.~\cite{greeks} to which we turn next.

\subsection{Low values of $\phi$~: Influence of ${\mathcal O}_h$}
\label{sec:IIIC}

\begin{figure}
 \begin{minipage}[t]{8cm}
 \centering
 \includegraphics[width=8cm,height=7cm]{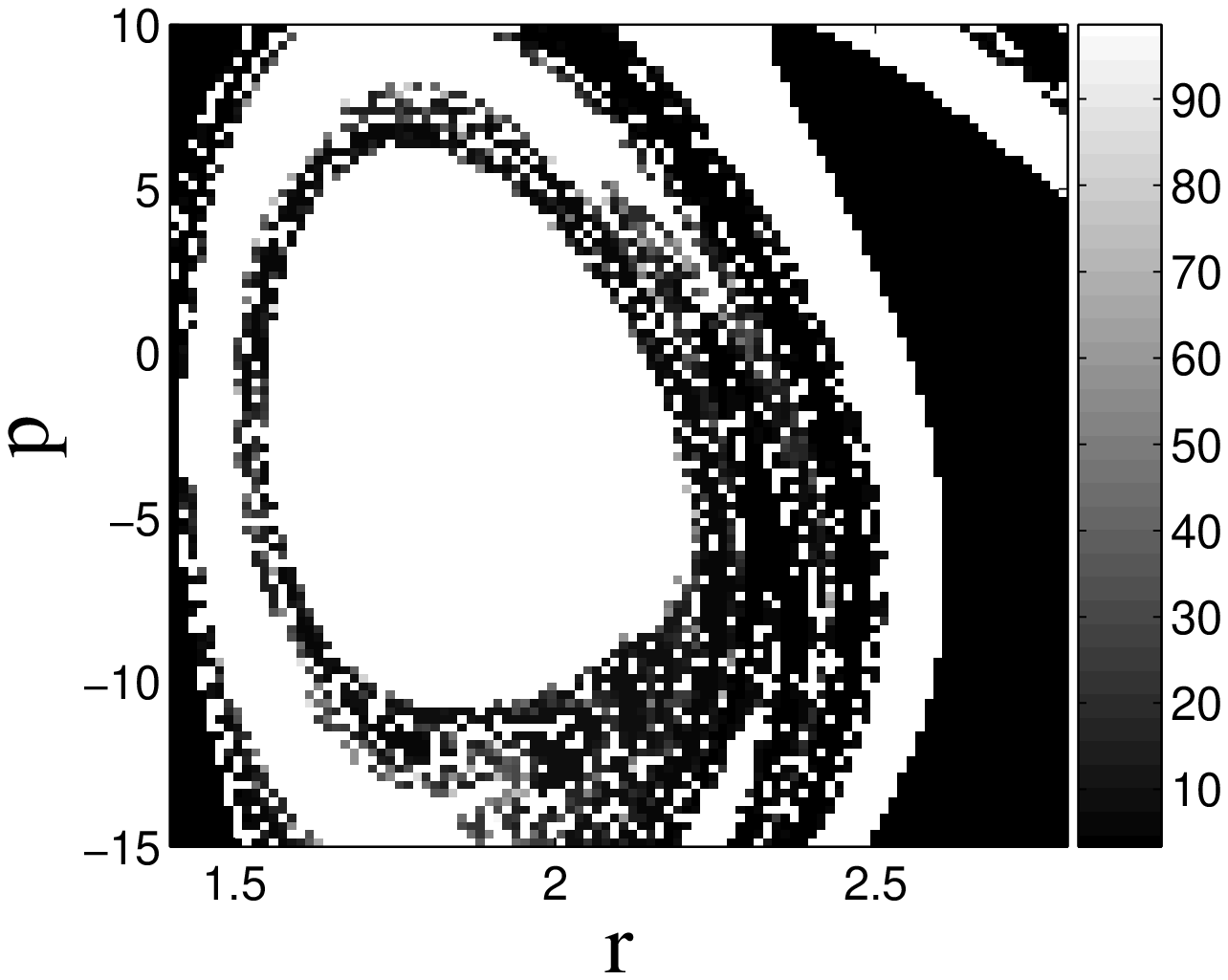}
 \mbox{{\bf (a)} }
 \end{minipage}
 \centering
 \begin{minipage}[t]{8cm}
 \includegraphics[width=8cm,height=7cm]{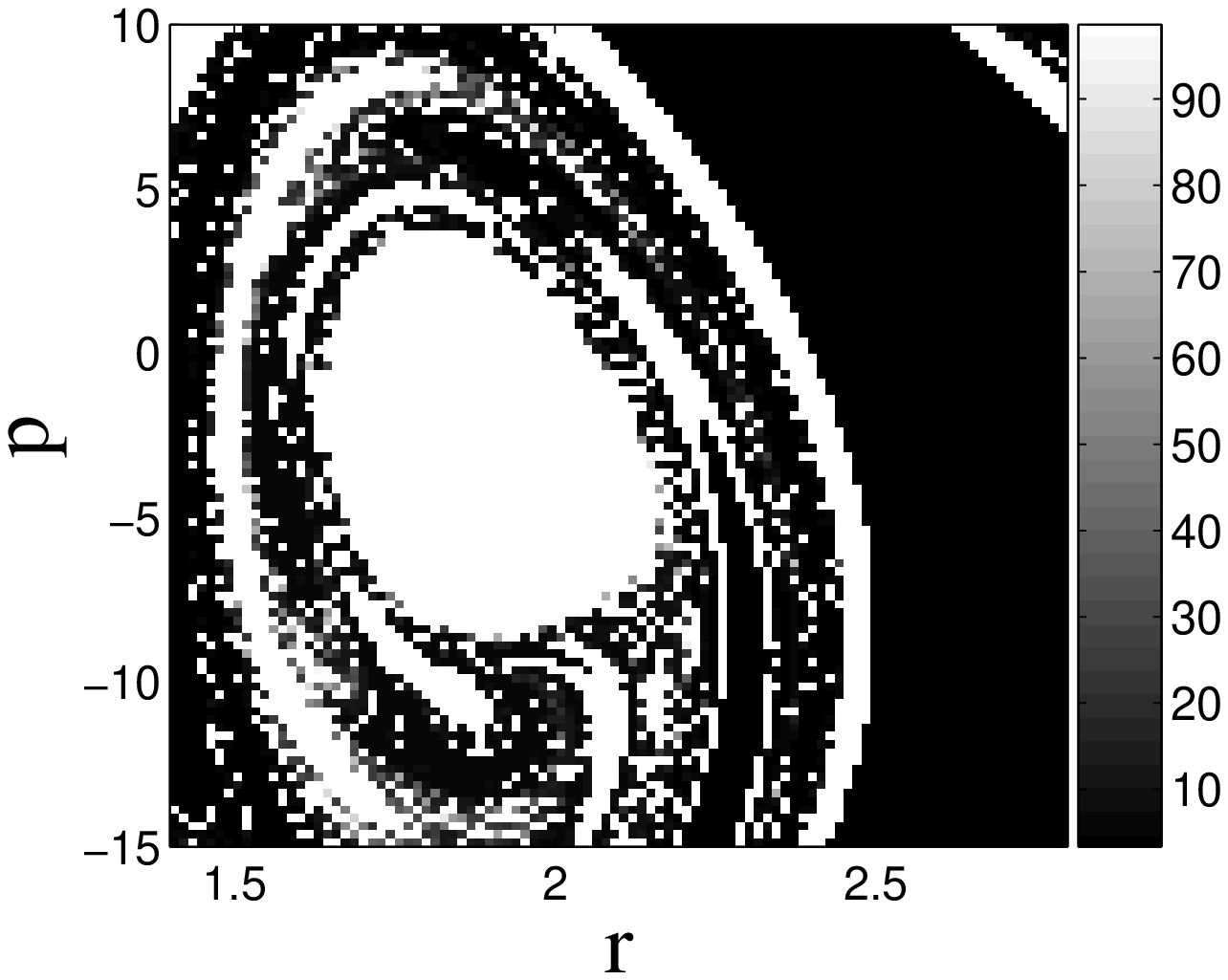}
 \mbox{{\bf (b)} }
 \end{minipage}
 \begin{minipage}[t]{8cm}
 \includegraphics[width=8cm,height=7cm]{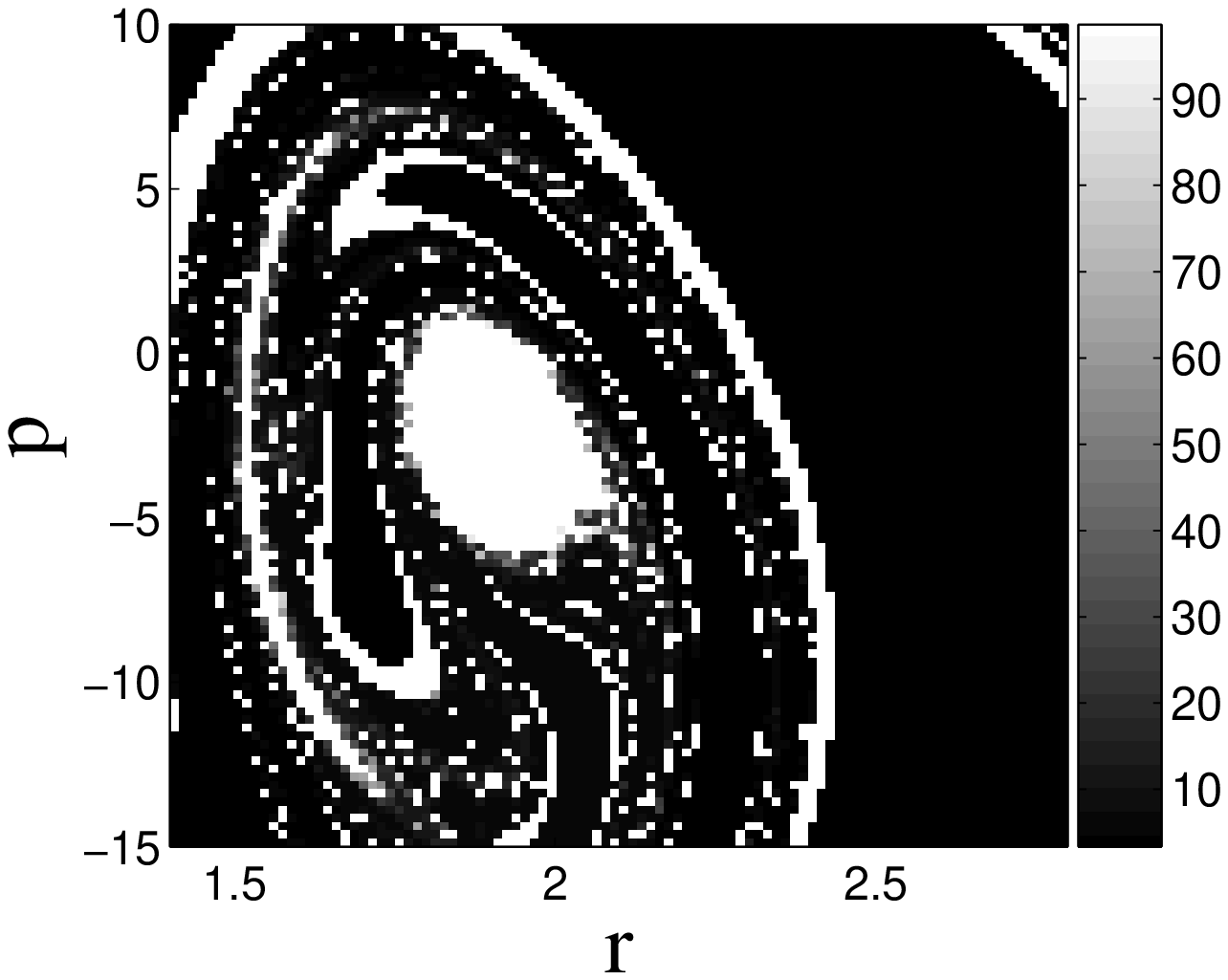}
 \mbox{{\bf (c)} }
 \end{minipage}
 \caption {\label{fig:laminaronline} Laminar plots with $\phi=0$ for $(a)$
$(F_1,F_2)=(0.1405,0.025)$, $(b)$ $(F_1,F_2)=(0.1559,0.04)$ and
$(c)$ $(F_1,F_2)=(0.165,0.055)$. The cutoff time is
$200\pi/\omega_{1}\approx 2244$ and the energy threshold is
$E_{th}=2$.}
\end{figure}

\begin{figure*}[p]
 \begin{minipage}[t]{8cm}
 \centering
 \includegraphics[width=8cm,height=7cm]{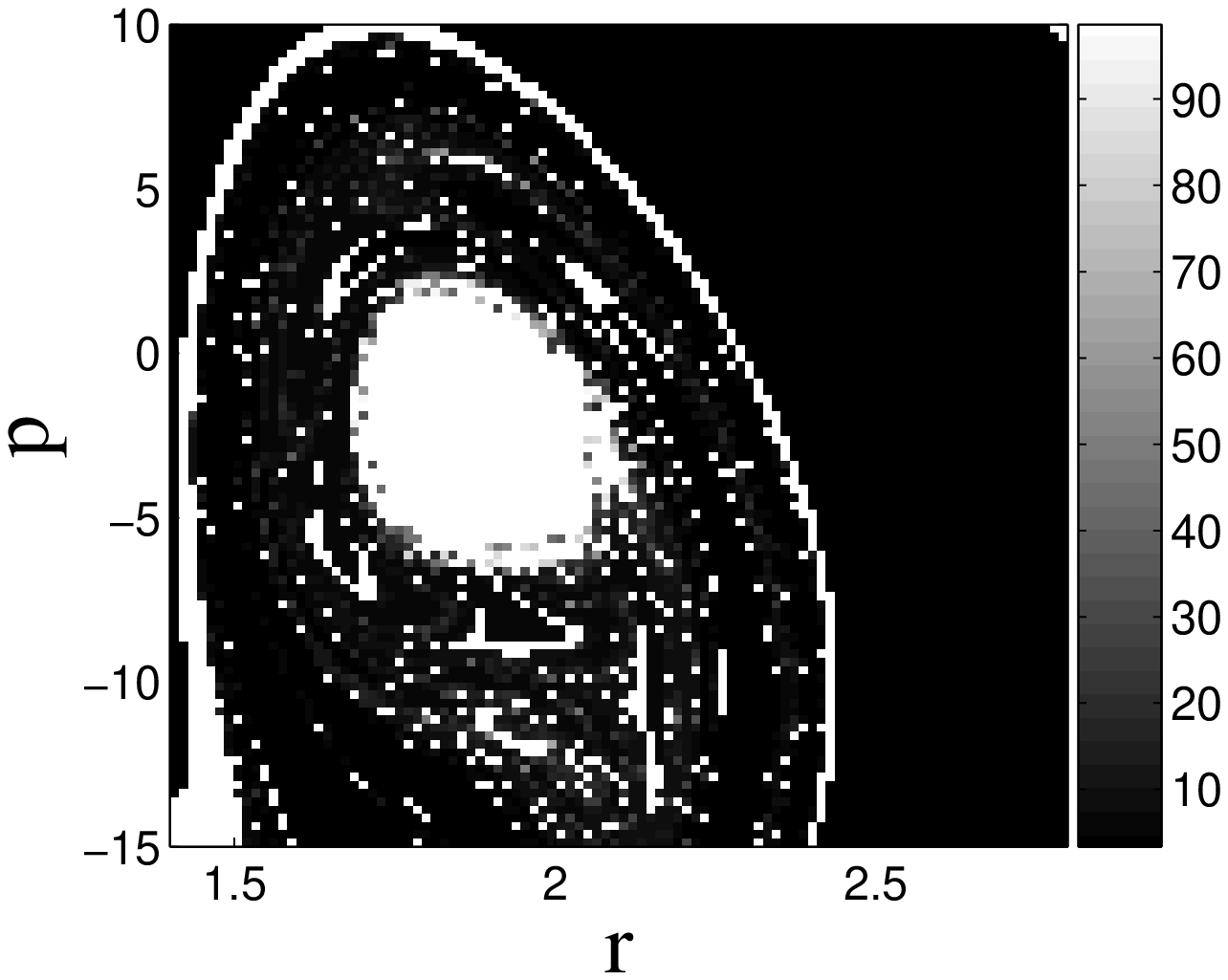}
 \mbox{{\bf (a)} }
 \end{minipage}
 \begin{minipage}[t]{8cm}
 \centering
 \includegraphics[width=8cm,height=7cm]{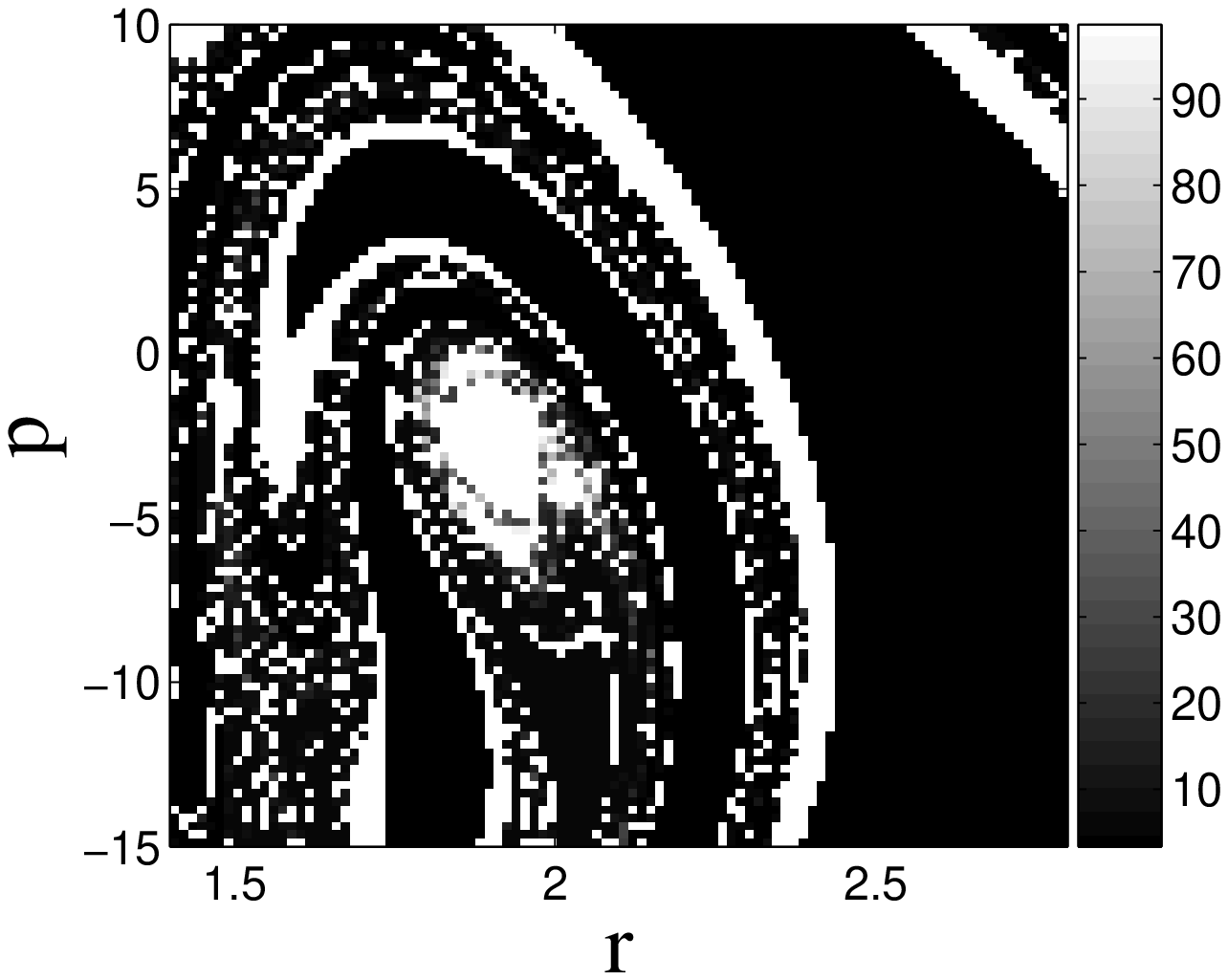}
 \mbox{{\bf (b)} }
 \end{minipage}
 \caption {\label{fig:laminarlocal} Laminar plots with $\phi=0$ for $(a)$
$(F_1,F_2)=(0.19,0.055)$, and $(b)$
$(F_1,F_2)=(0.14,0.055)$. The cutoff time is
$200\pi/\omega_{1}\approx 2244$ and the energy threshold is
$E_{th}=2$.}
\end{figure*}

For $\phi=0$, we observe two branches on Fig.~\ref{fig:fig2} $(a)$
where the residues of ${\mathcal O}_e$ vanish. Along these lines
we expect, from a linear stability analysis, locally a
constant degree of chaos, and hence dissociation
probability. However, this is not seen in direct
simulations since the dissociation probability actually increases along
these lines as $F_2$ is increased. This feature is shown
using laminar plots which represent contour plots of the number of
return times on the Poincar\'e section before dissociation
(defined as trajectories for which $E$ becomes greater than
$E_{th}=2$). The maximum integration time is
$200\pi/\omega_{1}\approx 2244$. Figure~\ref{fig:laminaronline}
shows laminar plots for the set of parameters on the line of
vanishing residues (as $F_2$ is increased) as marked by crosses in Fig.~\ref{fig:fig2}
$(a)$, and Figure~\ref{fig:laminarlocal} represents two additional
laminar plots for parameter sets transverse to that line (with a fixed $F_2$ and increasing $F_1$), marked by circles in Fig.~\ref{fig:fig2} $(a)$. These
plots show clearly that dissociation increases as $F_2$ is
increased. Even locally around the considered periodic orbit
${\mathcal O}_e$, there seems to be more chaos as $F_2$ is
increased. The situation is more complex at parameter values off
the vanishing-residue line as $F_1$ is varied on
Fig.~\ref{fig:laminarlocal} where the overall amount of
dissociation seems to be similar but with very different
distributions of dissociating trajectories.  All these features originate from the nonlinear
stability which can be captured by considering the linear
stability of higher order periodic orbits around the boundary of
the elliptic island ${\mathcal O}_e$.

\begin{figure}
 \begin{minipage}[t]{8cm}
 \centering
 \includegraphics[width=8cm,height=7cm]{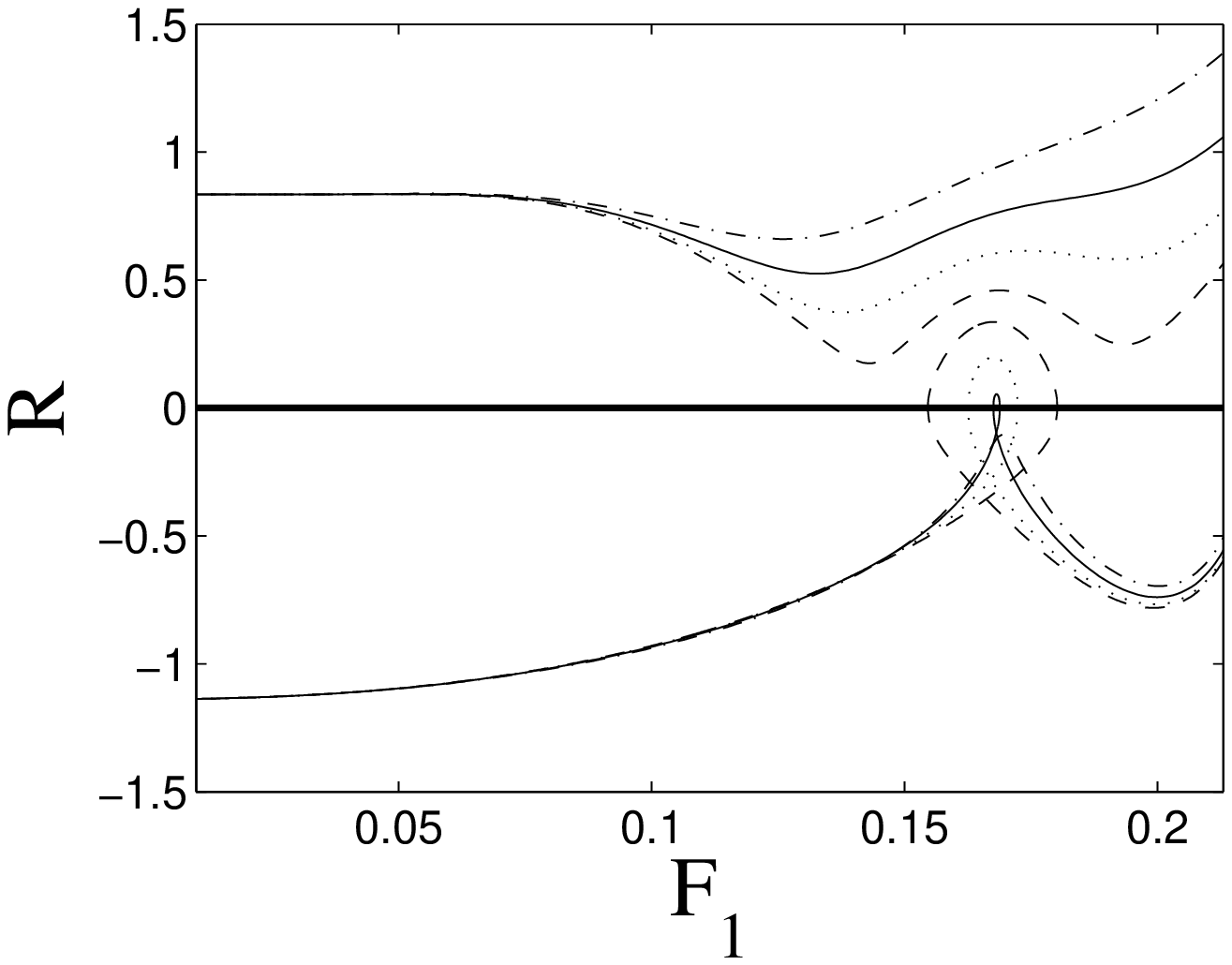}
  \mbox{{\bf (a)} }
 \end{minipage}
 \centering
  \begin{minipage}[t]{8cm}
 \includegraphics[width=8cm,height=7cm]{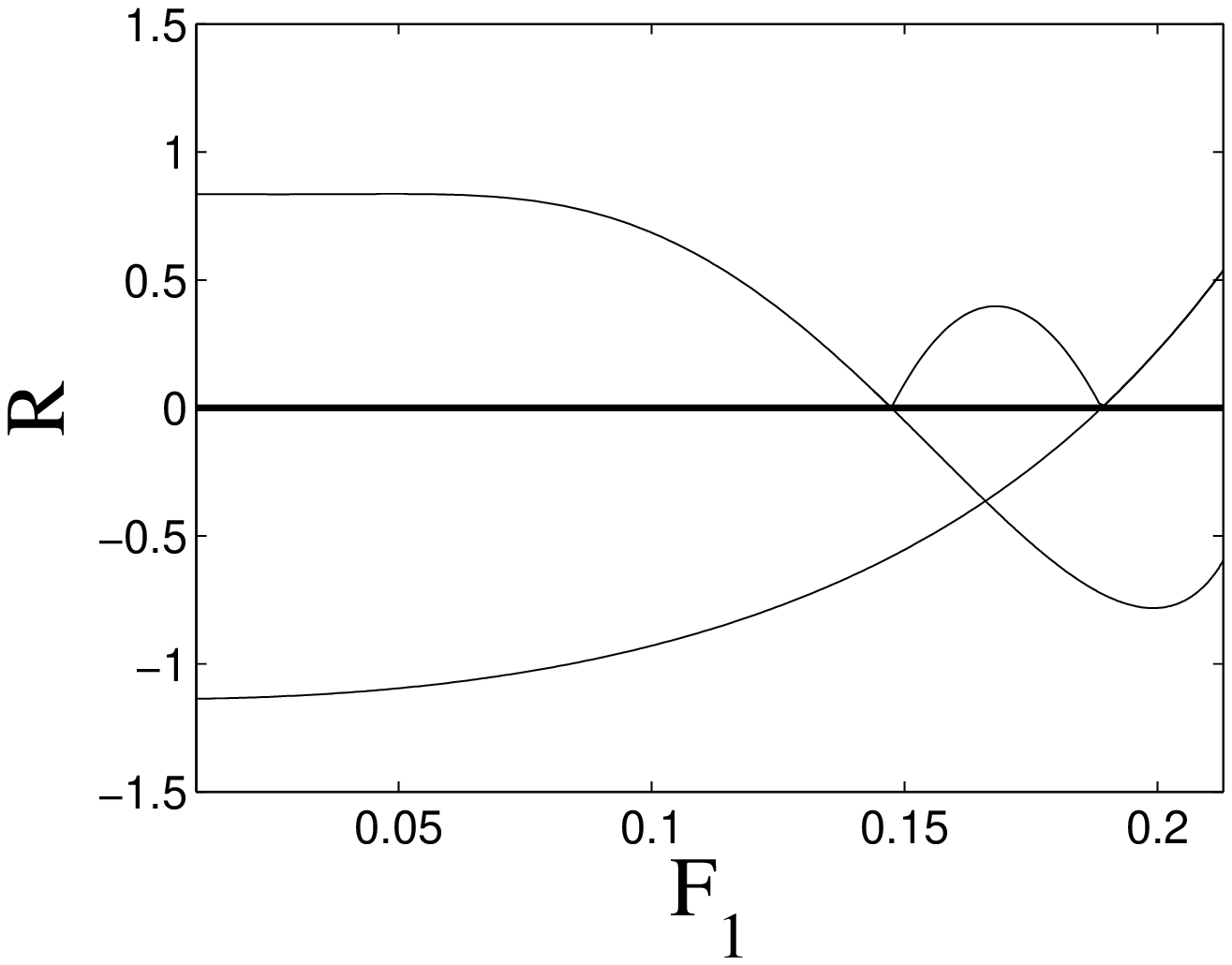}
 \mbox{{\bf (b)} }
 \end{minipage}
\caption{\label{fig:fig4} Residue versus $F_{1}$ for both elliptic
and hyperbolic periodic orbits at $F_2=0.03$ for $(a)$ several
different values of $\phi$~: $\phi=\pi/6$ (dash-dotted),
$\phi=0.35$ (solid), $\phi=0.2$ (dotted) and $\phi=0.06$ (dashed),
and $(b)$ $\phi=0$.}
\end{figure}

For insights into the parameter region where this
discrepancy occurs, we describe here the associated bifurcation.
In Fig.~\ref{fig:fig4} the residue as a function of $F_1$ is
plotted for both ${\mathcal O}_e$ and ${\mathcal O}_h$ with fixed
$F_2$ and $\phi$. We see clearly that there is a loop around
$R=0$, indicating that ${\mathcal O}_h$ undergoes a
bifurcation at $R=0$ which involves three periodic orbits of the
same period, two hyperbolic ones and an elliptic one. When $\phi$
is equal to $0$, the upper part of the loop merges with the
residue curve of ${\mathcal O}_e$, as shown in
Fig.~\ref{fig:fig4}$(b)$, for which the loop size reaches its
maximum size. The following picture emerges~: Without a loop in
the residue curve, the system has two periodic orbits with the
period of the field, ${\mathcal O}_e$ and ${\mathcal O}_h$. In the
region of parameters where there is a loop in the residues, the
system has four of these periodic orbits, two elliptic ones (which
are close to each other or even coincide at $\phi=0$) and two
hyperbolic ones. This additional hyperbolicity increases chaos (and hence dissociation)
locally.

\section*{Conclusions}

We have analyzed the dissociation dynamics of a model
diatomic molecule driven by a bichromatic field in terms of
periodic orbit bifurcations. Following the linear stability of a
few selected periodic orbits, we reproduced the dissociation
probability qualitatively in parameter space (two field amplitudes
and one relative phase). For relatively low $\phi$ and high
amplitudes $F_i$, the original hyperbolic periodic orbit
${\mathcal O}_h$ undergoes a particular bifurcation, which leads
to two branch lines on the $F_1$-$F_2$ residue plane. Along these two
branch lines, there is a discrepancy between predictions based
on the residues and direct simulations. The role of additional
periodic orbits is underlined regardless of whether the
discrepancy originates from bifurcated orbits (and the resulted
increase of hyperbolicity) or from higher-order periodic ones.

\section*{Acknowledgments}
This research was partially supported by the US National Science
Foundation. C.C. acknowledges support from Euratom-CEA (contract
EUR~344-88-1~FUA~F).

\listoffigures
\end{document}